\documentstyle[twoside,fleqn,espcrc2,epsfig]{article}

\title{Correlations in fluctuating geometries}

\author{P.~Bialas\address{%
Universiteit van Amsterdam, Instituut voor Theoretische
Fysica,\\ 
Valckenierstraat 65, 1018 XE Amsterdam, The
Netherlands}%
\thanks{ Permanent address: Institute of Comp. Science,
Jagellonian University, ul. Nawojki 11, 30-072 Krak\'ow, Poland}}

\begin{document}

\begin{abstract}
We compare two definitions of connected correlation
functions in fluctuating geometries. We show results of the MC simulations
for 4D dynamical triangulation in the elongated phase and compare them with
the exact calculations of correlation functions in the branched polymer
model. 
\end{abstract}

\maketitle

\section{Introduction}

Models of fluctuating(or random) geometries provide many challenging
problems.  One of them is the definition and interpretation of
correlation functions.  The usual formulation of the correlation
function as a correlator of two observables in two fixed points at some
distance apart is possible only if we have a fixed system of
coordinates and a metric. This may be suitable for the
perturbation thery approach. For theories like simplicial gravity,
formulated in coordinate independent way, this is impossible.  One way
to proceed is to sum  over all pairs of points with the fixed
distance between them:
\begin{eqnarray}\label{cor} 
G^{OP}(r)=<\!\sum_{ij}v_iv_jO_iP_j\delta_{d(i,j),r}\!>.
\end{eqnarray}
The above average is taken over all geometries and the sum runs over
all pairs of points at the fixed distance $d(i,j)=r$. The $v_i$'s
denote some suitable volume elements. Next we define a ``point-point''
correlator:
\begin{eqnarray}\label{concor}
<\!OP(r)\!>=\frac{G^{OP}(r)}{G^{11}(r)}
\end{eqnarray}
One immediate problem with those
definitions is that these are not strictly speaking a two--point
functions. The distance $d(i,j)$ depends on the geometry contrary to the
usual fixed lattice case and it cannot be pulled out of the average.
This leaves us with the average of a very non-local object.

\section{Connected correlation functions}

The next problem is the definition of the connected correlation functions.
The straightforward standard  substraction  fails as was shown in \cite{bs}.
The authors of \cite{bs} propose instead:
\begin{eqnarray}
\lefteqn{<\!OO(r)\!>_c^*=}\nonumber\\
&&\frac{1}{G^{11}(r)}
<\!\sum_{ij}v_iv_j(O_i-<\!O(r)\!>)\cdot\nonumber\\
\label{def1}
&&\cdot(O_j-<\!O(r)\!>)\delta_{d(i,j),r}\!>
\nonumber\\ 
&=&<\!OO(r)\!>-<\!O(r)\!>^2
\end{eqnarray}
where
\begin{eqnarray}
<\!O(r)\!>=\frac{G^{1O}(r)}{G^{11}(r)}
\end{eqnarray}
is the average of the observable $O$ over the spherical shells of
radius $r$. This definition has  nice properties, namely it vanishes
with increasing $r$ as we would expect for a correlation fuction. 
But it does not integrate to a susceptibility. A definition
which does integrate to  a susceptibility was proposed in \cite{bbkp}:
\begin{eqnarray}
\lefteqn{<\!OO(r)\!>^{**}_c=}\nonumber\\
&&\frac{1}{G^{11}(r)}
<\!\sum_{ij}v_iv_j(O_i-<\!O\!>)\cdot\nonumber\\
&&\cdot(O_j-<\!O\!>)\delta_{d(i,j),r}\!>
\nonumber\\ 
\label{def2}
&=&<\!OO(r)\!>-2<\!O\!><\!O(r)\!>+<\!O\!>^2
\end{eqnarray}
where $<\!O\!>$ is the usual average of $O$ over all the points of the lattice.

\begin{figure}[t]
\begin{center}
\epsfig{file=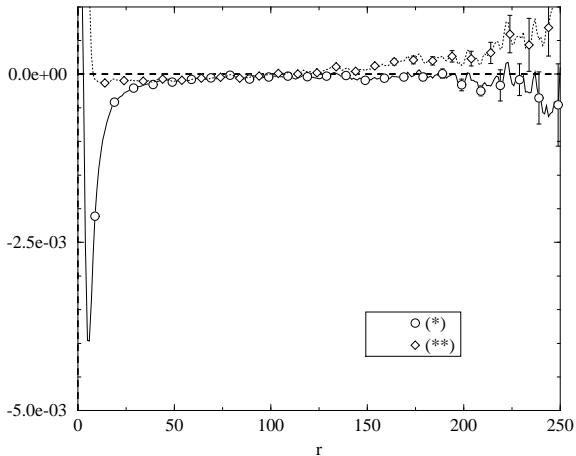,width=8.0cm,bbllx=50,bblly=70,bburx=552,bbury=430}
\end{center}
\caption{\label{f1}Correlation functions $<\!\!RR(r)\!\!>^{*}_c$  and $<\!\!RR(r)\!\!>^{**}_c$ in  4D simplicial gravity, $\kappa_2=1.5$, 32k simplices.
For clarity only one in ten points is plotted.}
\end{figure}

In the case of the fixed geometry $<\!O(r)\!>=\\<\!O\!>$ and both
definitions give indentical results. This is not so for fluctuating
geometries.  We plotted the two definitions with the observable being
Regge curvature in simplicial 4D gravity in figure~\ref{f1}.  They
differ quite distinctly. The short distance behavior it totally
different.  For the large distances one can see that the definition
(\ref{def2}) begins to deviate from zero, and as we will show later
this is not due to errors but it is an intrinsic property of this
function.

\subsection{Probabilistic interpretation}

As to get some idea as to what causes those differences we can look at
those function from the point of view of probability theory.

Let's suppose that we make list of all pairs of points separated by
the  distance $r$ on all configurations. To each pair we
assign a weight equal to the weight of the configuration times the volume
elements of each point.  We pick up a pair at random from this
weighted list. The values of observable $O$ at each point of the pair
will then define two random variables: $O_1$ and $O_2$.  
Then 
\begin{eqnarray}
<OO(r)>^{*}_c=E(O_1O_2)-E(O_1)E(O_2)
\end{eqnarray}
where $E(O)$ denotes the expectation value of the operator $O$.  The
expression on the right hand side of this equation is zero if and only
if the two variables are independent, so the definition (\ref{def1})
provides a direct measure of the degree of correlation between 
the values of observable $O$ in points at the distance $r$ apart.

To interpret the definition (\ref{def2}) we have to proceed a little
differently. We make up a list of all points on all configurations.
To every point we asign the weight of the configuration times its
volume element. We pick up the random point $i$ from the list.  We
define now two random variables: $O_i$ the value of observable $O$ at the
point and $S_i=\sum_{j}v_j\delta_{d(i,j),r}$ the volume of the
spherical shell of radius $r$ around this point. Then
\begin{eqnarray}
\lefteqn{<\!O(r)\!>-<\!O\!>=}\nonumber\\\label{pb2}
&&\frac{V}{G^{11}(r)}
(E(O S)-E(O)E(S))
\end{eqnarray}
It easy to check that the square of the expresion on the left-hand
side of the above equation is the difference between the definitions
(\ref{def1}) and (\ref{def2}).  So the definition (\ref{def2}) mixes
in also the correlations between the value of an observable and the
volume of spherical shell around it.

\section{Branched polymers}

To study these issues in more detail than permitted by todays status of
the simplicial gravity simulations one can use other models of random
geometry. One of such models is the branched polymer\cite{adfo}. 
This model can be solved and correlation functions can be calculated for
operators $q_i$ and $\log q_i$, where $q_i$ is the number  of branches
of vertex $i$ \cite{b}. Interesting features of this particular model are that
it describes the elongated phase of 4D simplicial gravity \cite{aj}, and that it exibits a phase transition between a ``short'' and a ``long'' phase
reminiscent of the transition in 4d simplicial gravity\cite{bb}.

The results from this model for the function (\ref{def1}) are the
following:\\
 i) In the grand canonical ensemble this function is zero
for positve values of $r$. This in a sense is a defining feature of the model:
the branching probabilities are independent.\\
ii)
In the canonical ensemble this function in the thermodynamical limit
is power--like:
\begin{eqnarray}
<\!\log(q)\log(q)(r)\!>^{*}_c\propto\frac{1}{(a+r)^2}
\end{eqnarray}
The correlations appear because the number of vertices is fixed.
This is to be expected, but what is unusual is that those correlation
persist even when the number of vertices grows to infinity. A possible
mechanism could be the following: if we know that a vertex has one
branch then its unique neighbour must have more than one branch, this
introduces a correlation between nearest neighbour vertices which does
not depend on the number of vertices.  We believe that this effect is
propagated to longer distances.\\ 
iii) Finite size effects alter this
behaviour with the net effect of flattening the function at large $r$.

\begin{figure}[t]
\begin{center}
\epsfig{file=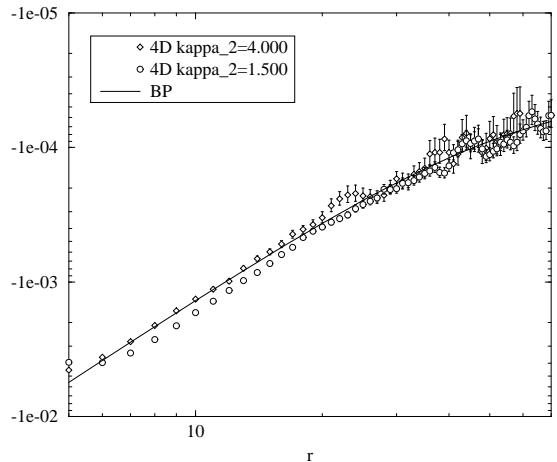,width=8.0cm,bbllx=50,bblly=70,bburx=552,bbury=430}
\end{center}
\caption{\label{f2}Correlation function $<\!\!RR(r)\!\!>_c^{*}$ for 4d simplicial gravity(32 k simplices) compared to the $<\!\!\log(q)\log(q)(r)\!\!>_c^{*}$ correlation function
of branched polymers}
\end{figure}

In figure \ref{f2} we have ploted the results of MC simulations in the
elongated phase of 4d simplicial gravity. The effects described above
are clearly visible: the functions are power like at short
distances (with power exponent equal to 2.0 within errors) and then
flatten due to the finite size with very good accord to the BP
predictions.

For the definition (\ref{def2}) the results from BP are:\\ 
i) This
function is not zero in the grand canonical ensemble.  Althought the
formula (\ref{pb2}) is not valid in this ensemble a similar
interpretation exists. This shows that this function contains also the
correlations between the number of vertices at some distance from the
point, and the number of branches of this point. Those are clearly
dependent even in the grand canonical ensemble. \\ 
ii) In the
canonical ensemble this definition is zero in thermodynamical limit
($r>0$).  This implies a strong relation between branch-branch and
branch-volume correlation, which is probably specific to the BP
model. \\ 
iii) The convergence to the thermodynamical limit is not
uniform.  In particular the function will tend away from zero for
$r\gg1$ for any finite volume.
\begin{figure}[t]
\begin{center}
\epsfig{file=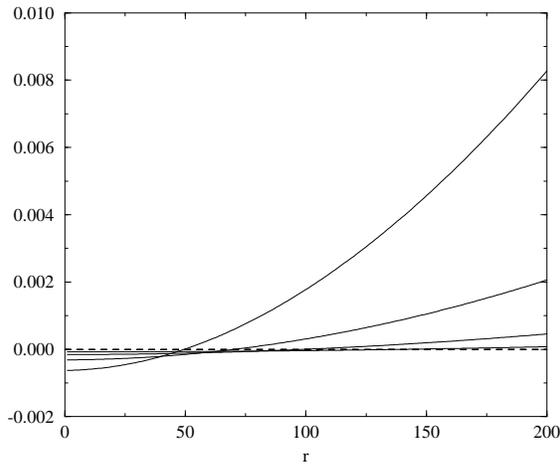,width=8.0cm,bbllx=50,bblly=70,bburx=552,bbury=430}
\end{center}
\caption{\label{f3}Correlation functions $<\!\!\log(q)\log(q)(r)\!\!>^{**}_c$ for branched polymers with two, four, eight and 16 thousand vertices. As the size of the polymer grows, the curves approach to zero.}
\end{figure}

These results are illustrated in the figure~\ref{f3}. It shows the
$<\!\log(q)\log(q)(r)\!>^{**}_c$ correlation functions for branched
polymers of various sizes. As the size increases the functions go to
zero for any fixed $r$. However for each fixed size if we increase $r$
the functions will eventually go away from zero. We also see that this
behavior agrees qualitatively with the results of MC simulations  shown
on figure~\ref{f1}.

\section{Discussion}

We have compared two definitions of connected correlation functions on
the fluctuating geometries. Both were introduced in the context of
simplicial gravity, each with a different goal in mind.
To really compare and understand them we need a more profound
understanding of the theory that we have now. One way to proceed is to
study  finite size scaling. This was the way proposed in
\cite{bbkp}.  The other way is to study the interaction of particles in
the theory, this is in the spirit of \cite{bs,bs2,s}.  It may happen
that the functions describing the finite size scaling and the
interaction potential are different.  Both  goals are still to be
attained, and are quite difficult to pursue  due to the enormous time
required by simulations and very few theorethical results. That is why
we propose the BP model as a very promising tool for gaining insight
into these issues.
As shown above this model provides an excellent description of the
elongated phase of simplicial 4D gravity. In view of results from 
\cite{bb} it can hopefully 
also provide information about the critical region.

\section*{Acknowledgments} 
I would like to thank Z.~Burda, B.~Petersson
and J.~Smit for many helpful comments and discussions. The simulations
were done on the SP2 computer at SARA. This work was supported
by Stichting voor Fundamenteel Onderzoek 
der Materie (FOM) and partially by KBN (grant 2P03B 196 02).

\end{document}